\documentclass{jfm}
\usepackage{graphicx}
\usepackage{amssymb,amssymb}
\usepackage{latexsym}    

\begin{document} 

\title[Crucial role of side walls for granular surface flows: 
consequences for the rheology]
{Crucial role of side walls for granular surface flows: 
consequences for the rheology}

\author[Pierre Jop, Yo\"el Forterre and Olivier Pouliquen] 
{P\ls I\ls E\ls R\ls R\ls E\ns  J\ls O\ls P\ls , \ns   Y\ls O\ls \"E\ls L \ns F\ls O\ls R\ls T\ls E\ls R\ls R\ls E\ns 
\and O\ls L\ls I\ls V\ls I\ls E\ls R \ns  P\ls O\ls
U\ls L\ls I\ls Q\ls U\ls E\ls N }

\affiliation{IUSTI, Universit\'e de Provence, CNRS UMR 6595,\\ 5 rue 
Enrico Fermi, 13453 Marseille cedex 13, France.} 

\date{\today} 

\maketitle

\begin{abstract}
In this paper we study the steady uniform flows that develop when granular material is released from a hopper on top of a static pile in a channel. We more specifically focus on the role of side walls by carrying out experiments  in setup of different widths, from  narrow channels 20 particle diameters wide to channels 600 particle diameters wide. Results show that steady flows on pile are entirely controlled by side wall effects. A theoretical model, taking into account the wall friction and based on a simple local constitutive law recently proposed for other granular flow configurations (\cite{gdrmidi04}),  gives predictions in quantitative agreement with the measurements. This result gives new insights in our understanding of free surface granular flows and strongly supports the relevance of the  constitutive law proposed. 

\end{abstract}

\section{Introduction}

\begin{figure}
	\begin{center}
\includegraphics[scale=0.7]{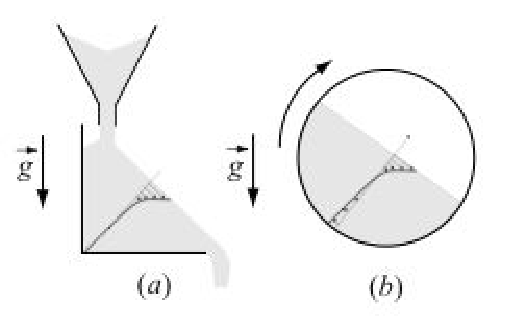}
\caption{The two configurations leading to granular surface flows. (\emph{a}) Flow on a heap; (\textit{b}) Rotating drum.}
 \label{fig:drumpile}
 \end{center}
\end{figure}

Grains avalanching on a sand heap are often presented as the archetype of granular flows. In this situation both the liquid and solid behaviours of granular material coexist. The flow is confined in a thin layer of  grains, typically   few grain diameters, flowing at the free surface of a pile formed by static grains.  Despite the apparent simplicity of the configuration, its description still represents a serious challenge as it gathers in a single flow all the difficulties specific to granular material. How can we describe the transition between the liquid  and the solid behaviour? How is the thickness of the flowing layer selected? What kind of constitutive laws can describe the flow characteristics?
These questions have motivated many experimental works in the ten last years. Two different configurations have been used to create  surface granular flows: the flow on a heap and the rotating drum (figure \ref{fig:drumpile}). In the first one, the granular material is released  from a hopper on top of a pile. The control parameter is the flow rate.  The second one is a cylinder half filled with the granular material. In this case, the control parameter is the rotation speed that indirectly controls the flow rate of the flowing layer. In both configurations, the flow characteristics are similar and can be summarized as follow:

(i) A minimum flow rate (or rotation rate in the rotating drum) exists below which the flow is intermittent and proceeds as successive avalanches. Above this threshold, a steady regime is observed (\cite{rajchenbach90}; \cite{lemieux00}).

(ii) In the steady regime, the free surface inclination, sometimes called the dynamical angle of friction, increases when increasing the flow rate (\cite{khakhar01}; \cite{ancey01};  \cite{taberlet03}; \cite{gdrmidi04}).

(iii) In the steady regime, the velocity profile (often measured at the side wall) is localized at the free surface, with a linear profile at the top followed by an exponential tail (\cite{komatsu01}; \cite{gdrmidi04}). The shear rate in the flowing region is of order  $\sqrt{g/d}$ where $g$ is the gravity and $d$ the particle diameter, this shear rate being weakly dependent on the flow rate (\cite{rajchenbach90}; \cite{khakhar01}; \cite{bonamy02a}).

Based on these observations, several theoretical approaches have been proposed.
Some are based on the analysis of individual grains motion, showing that a linear profile localized at the free surface could be explained by the dissipative nature of collisions (\cite{andreotti01}; \cite{rajchenbach03}; \cite{hill03}).  Others are hydrodynamic descriptions based on different rheological constitutive laws (\cite{elperin98}; \cite{khakhar01}; \cite{bonamy02a}; \cite{josserand04}).  In all these attempts to describe free surface flows, the influence of side walls is neglected. The experimental measurements are interpreted as if the side walls, often made of smooth surfaces, do not play a major role in the flow dynamics.

However, this assumption is far from being straightforward and to our opinion warrants a deeper investigation. Some signs exist in the literature showing that side walls could have a crucial role in the dynamics of surface granular flows. The most striking evidence concerns the inclination angle of the free surface with respect to horizontal. In the intermittent avalanching regime, experimental works have shown that angles at which avalanches start or stop strongly depend on the width between lateral walls (\cite{grasselli97}; \cite{zhou02}). In the continuous flowing regime, experiments carried out in long rotating drums have shown that the surface inclination is few degrees higher close to the wall than in the middle of the drum (\cite{dury98}; \cite{yamane98}). More recently, some studies of flow on a heap have revealed the existence of surprisingly steep piles at high flow rates in narrow channels (\cite{taberlet03}).   This angle made by the  free surface with horizontal controls the stress distribution in the material. As a consequence, if side walls have a strong influence on the inclination, it is legitimate to wonder to what extent they do not dramatically also modify the flow characteristics, such as the flow thickness and the velocity profiles. However, most of the experiments are carried out in narrow devices, with a gap between side walls less than 30 particle diameters. Very few studies are made in wide rotating drums configuration and none in wide heap configuration. What does happen in the case of large system? Are flow characteristics going to change? Does the influence of side walls disappear for large enough system?

Answering these questions on the role of side walls for granular surface flows represents the first goal of this study. Once we know if side walls can be neglected or not and what is their influence, it will be possible to properly use the surface flow configuration to test constitutive equations  for dense granular flows.  This is the second goal of this study, where we shall test an empirical rheology recently proposed for other configurations.

 In a recent collective work (\cite{gdrmidi04}), data from different configurations and from different research groups have been collected. A local rheology has been proposed, which was able to unify results numerically obtained on plane shear (\cite{chevoir04}; \cite{ iordanoff04}; \cite{dacruz04}) and results obtained both in experiments and in simulations for flows down inclined planes (\cite{pouliquen99a}; \cite{chevoir01}; \cite{pouliquen02b};  \cite{silbert03}; \cite{forterre03}).  The  rheology is written has a friction law, \emph{i.e.}\,\,the shear stress is proportional to the normal stress, with a coefficient of friction depending on a dimensionless shear rate.  Although this local approach suffers from several limits which are discussed in GDR MiDi (2004), the fact that a single law correctly describes two different configurations, the plane shear and the inclined plane, is to our opinion very encouraging and deserves a deeper confrontation with other flow configurations. In that sense, granular surface flows represent a severe test for this local rheology as they exhibit peculiar flow characteristics, \emph{e.g.}\,\,localized velocity profiles,  that strongly contrast with those observed in plane shear or inclined plane configurations.

The aims of this study are then, first to clarify the role of side walls for granular surface flows and second, to test the local constitutive law proposed recently in GDR MiDi (2004). The configuration we choose is the flow on a heap (figure \ref{fig:drumpile}a). This choice rather than the rotating drum configuration is motivated by the fact that steady uniform flows, \emph{i.e.}\,\,with a flowing thickness independent of the distance to the hopper, can be obtained for heap flows whereas in rotating drums the thickness varies along the flow. For this configuration we  systematically investigate the role of  side walls by making experiments in channels of different widths from a quasi 2-D configuration (20 particle diameters) to a very wide situation (600 particle diameters). We then compare our measurements of the free surface inclination, the free surface velocity and the flow thickness in the steady uniform regime to the prediction of the local rheology.

The paper is organized as follow. The experimental set-up and measurement methods are presented in section \ref{experiment}.  The influence of the channel width on the steady uniform flow properties  are presented in section \ref{sidewall}.  The theoretical approach based on the local rheology proposed recently in GDR MiDi (2004) is  described in section \ref{theory}, and the predictions are compared with measurements in section \ref{comparison}.  Discussion and conclusion are given in section \ref{discussion} and \ref{conclusion}.

\section{\label{experiment}Experimental  method}
 
 \subsection{Experimental set-up}
\begin{figure}
   \begin{center}
  \includegraphics[scale=0.8]{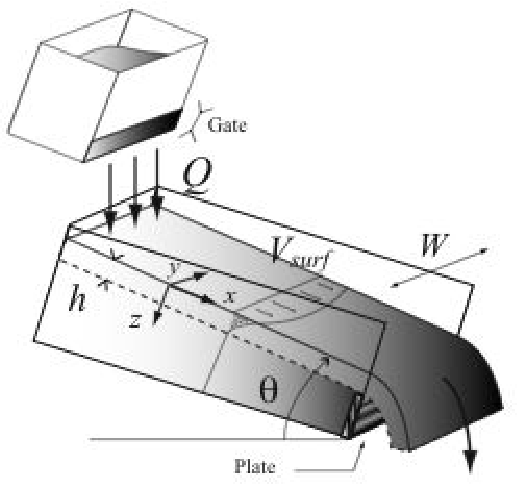}
  \caption{\label{fig:setup} Experimental setup.}
  \end{center}
 \end{figure}

 The experimental set-up is sketched in figure \ref{fig:setup}. We use glass beads of diameter $d=0.53$ $\pm 0.05$ mm as granular material. The box is 1.5 m long made of a rough  bottom plate  and  two smooth lateral glass walls. The gap $W$  between the side walls can be changed from 1 cm ($19 d$) up to 30 cm ($570 d$). The box is closed at the bottom end by a 10 cm height plate. In order to use less granular material, the whole set up is inclined at an angle less than the angle of repose of the material, so that, when the box is filled, the static layer is everywhere thicker than 10 cm ($190 d$). On top of this static layer we release particles from a hopper placed at the top of the channel. The flow rate is controlled by the opening of the gate. The experimental procedure is the following. At the beginning of each experiment the box is empty. When opening the gate of the  hopper, the beads first  fill the box before flowing over the end plate when the box is full. For high enough flow rates, a steady regime then develops with a   layer of grains flowing on top of a static pile. All the measurements presented in the following are done in the steady regime.
 
 In this regime, far from the entrance and the exit, the flow is uniform along the $x$-direction. The free surface make a constant angle $\theta$ with the horizontal and is almost flat in the transverse direction (its variations between the wall and the middle are less than 2 grain diameters). The free surface velocity is aligned along the $x$-direction, the transverse velocity being less than 1\% of the longitudinal one, and. The free surface velocity is also uniform along the $x$-direction however it varies in the transverse y-direction: the flow is faster in the middle of the channel. The thickness of the flowing layer also varies in the $y$-direction and is larger in the middle than at the walls.
In order to characterize the flow, four quantities are measured: the mean flow rate per unit of width $Q$ (m$^2$/s) (in the following we simply call $Q$ the ``flow rate"), the inclination of the free surface with horizontal $\theta$, the thickness of the flowing layer $h$ measured in the centre line of the channel, and the free surface velocity field $V_{surf}(y)$. 
It must be emphasized that, in this configuration, the only control parameter is the mean flow rate  $Q$ fixed by the aperture of the hopper. The thickness $h$, the velocity $V_{surf}$ and the inclination $\theta$ are chosen by the system. It should be noticed that the control parameter $Q$ is a flow rate averaged over the channel. The local flow rate is not uniform and is smaller close to the wall than in the middle of the flow.

 \subsection{Measurement methods}
 In order to measure $Q$, we weigh the mass $m$ flowing out of the channel during a finite time $T$ (between 30 s and 3 s) once the steady regime is reached.  $Q$ is then computed knowing the density of the beads $\rho_s=2450$ kg/m$^3$ and assuming the volume fraction of beads during the flow to be $\phi=0.6$: $Q=m/(TW\rho_s\phi)$. The inclination of the free surface $\theta$ is  measured from a picture taken from the side and analyzed using image processing (software ImageJ available from the internet at {\it http://rsb.info.nih.gov/ij/}). The measurement precision is about $\pm 0.04^o$. Velocity profiles are computed from movies recorded by a fast camera taken at rate between 250 and 1000 frames per second. The velocity profile is then obtained using a modified Particle Imaging Velocimetry (PIV) method. The precision is about $\pm 3$ mm/s. Finally we have to measure the flow thickness. 
 
\begin{figure}
  \begin{center}
  \includegraphics[scale=0.5]{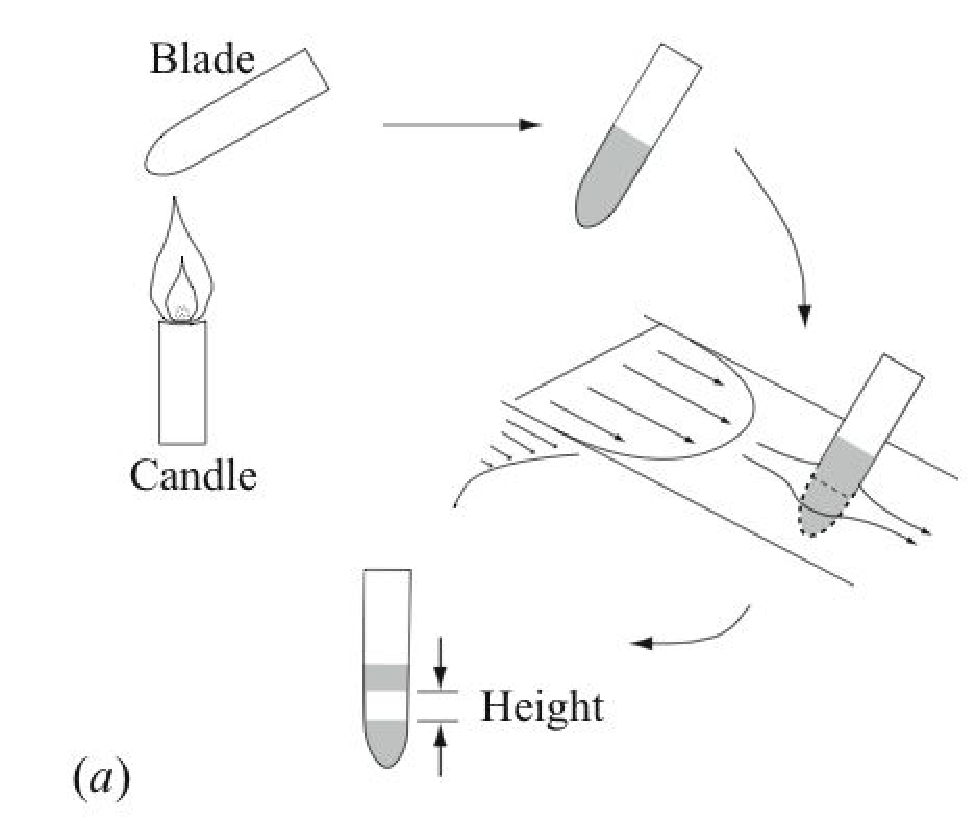}
  \includegraphics[scale=0.7]{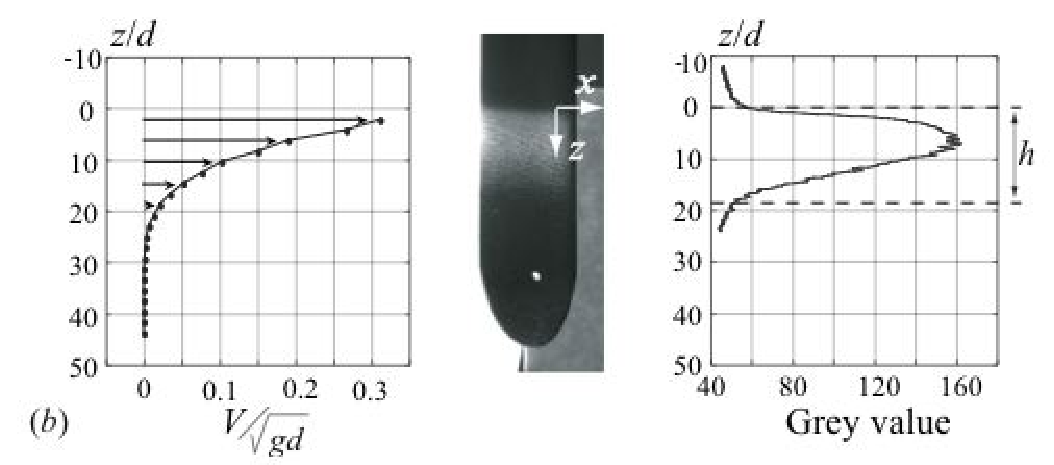}
  \caption{\label{fig:blade} (\textit{a}) Thickness measurement method. (\textit{b}) Calibration of the erosion method. left: Velocity profile at side wall. centre: Black soot on the blade eroded by the flow. right: Grey value along the blade. The flow thickness $h$ is estimated as follow: the top part of the curve is fitted by two straight lines and the bottom part is fitted by a modified Gaussian. $h$ is defined as the distance between the intersection of the lines and the point of maximum curvature of the fitted curve.}
 \end{center}
 \end{figure}

 Since the granular material is opaque, we cannot measure $h$ in the bulk using a camera. In order to get an estimate of the flow thickness far from the walls, we have developed a weakly intrusive method based on erosion measurements. A thin metal blade, $0.45$ mm thick and 2 cm wide, is blackened with the flame of a candle. The blade is then suddenly immersed in the flow, perpendicular to the free surface and hold at a fixed position during 20 s before we remove it  (figure \ref{fig:blade}\textit{a}). We have checked that the upstream flow is hardly perturbed by the intrusion of the thin blade. Once the blade is removed, one observes that the black soot has been eroded in the region where the grains were flowing but not in regions where grains were static. In order to test the method we have put the blade close to the wall, where we can compare the erosion pattern with the depth velocity profile along $z$ measured using PIV. A typical result is shown in figure \ref{fig:blade}(\textit{b}) where the centre picture shows the eroded blade lighted at a low incidence angle.  The bright part corresponds to region eroded by the flow. The graph on the right is the intensity profile of the blade averaged over the width. For comparison we have plotted on the same figure the corresponding velocity profile (figure \ref{fig:blade}\textit{b} left). The correlation of the erosion pattern and the velocity profile is good showing that the erosion method gives a good estimate of the flow thickness $h$ (see legend of figure \ref{fig:blade}).
 
\section{\label{sidewall}Influence of the side walls on steady uniform flows}

In order to precisely understand the influence of the side walls on granular surface flows, measurements are made for channel widths varying from 19 particle diameters to 570 particle diameters. For each width, the inclination $\theta$, the thickness $h$ and the free surface velocity $V_{surf}$ have been measured as a function of the flow rate $Q$.  The results presented in the following are expressed in terms of dimensionless variables using the particle diameter $d$ and the gravity $g$. The dimensionless flow rate is $Q^*=Q/d\sqrt{gd}$, the dimensionless thickness is $h^*=h/d$, the dimensionless width is $W^*=W/d$ and the dimensionless velocities are $V^*=V/\sqrt{gd}$.

\subsection{Existence of steady uniform flows}
\begin{figure}
    \begin{center}
  \includegraphics[scale=0.50]{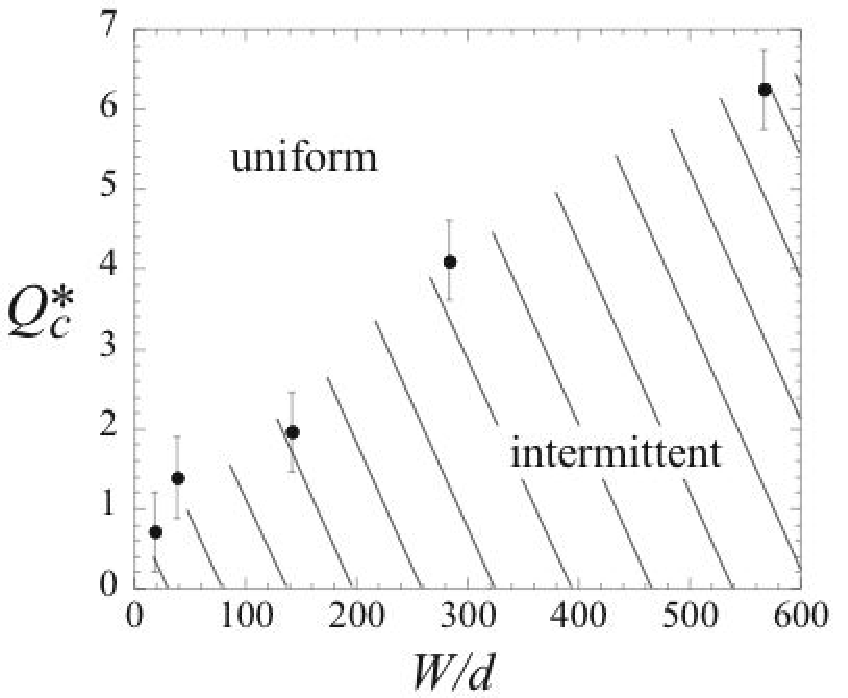}
 \caption{\label{fig:qc} The critical flow rate per unit of width, $Q^*_c$, separating the intermittent regime and the stationary flow as a function of $W/d$.}
  \end{center}
 \end{figure}

The first observation is that a minimum flow rate $Q_c$ is necessary to obtained a steady flow (\cite{rajchenbach90}; \cite{lemieux00}). Below this critical value, the flow is intermittent and occurs by successive avalanches.  Above this value, a steady regime is reached, where the outlet flow rate is constant and equal to the inlet flow rate. The interesting point we show here is that $Q_c$ depends on the width of the channel $W$ as shown in figure \ref{fig:qc}. For wide channels, one needs to supply more grains per second and per unit of width to reach a stationary flow.
    
In the stationary regime, the spatial variation of the flow along the slope is as follows.  The grains falling from the hopper first accelerate at the top of the pile, before reaching a steady velocity. About 10 cm from the outlet, they accelerate again and fall over the end plate. In between the entrance and the outlet regions,  the flow is then uniform independent of the $x$ position. All the measurements presented below are done in this steady and uniform regime. 

\subsection{Free surface inclination}

\begin{figure}
  \begin{center}
  \includegraphics[scale=0.6]{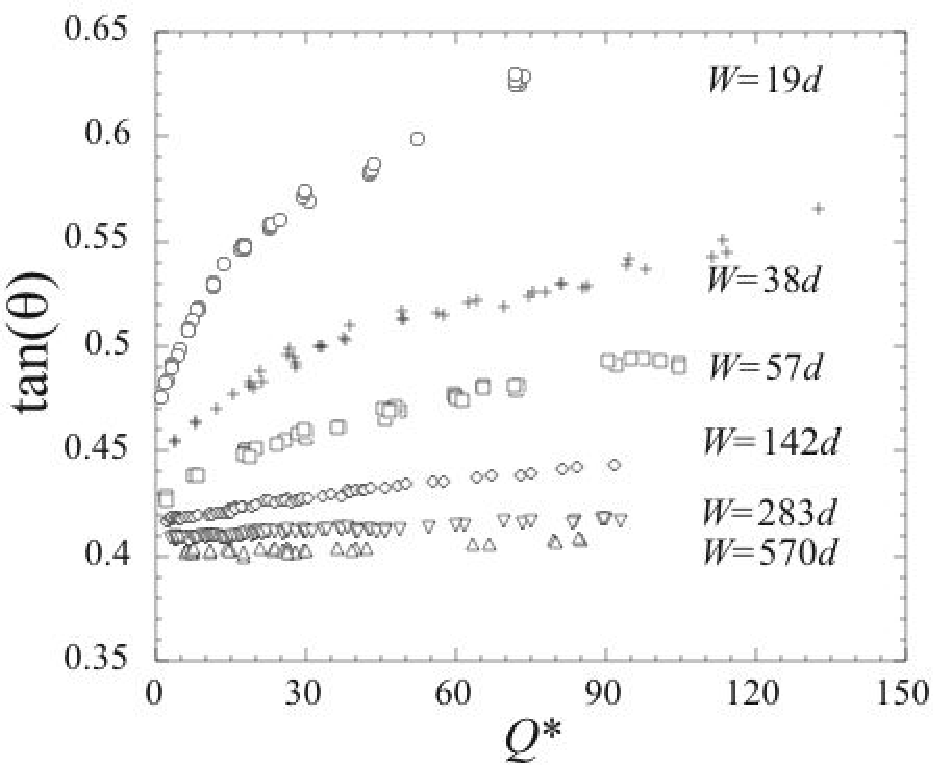}
   \caption{\label{fig:tan1} Surface slope $\theta$ as a function of the dimensionless flow rate $Q^*$ for different widths.}
  \end{center}
 \end{figure}

As observed in previous studies (\cite{rajchenbach90}; \cite{dury98}; \cite{yamane98}; \cite{grasselli99}; \cite{lemieux00}; \cite{khakhar01}), the angle between the free surface and the horizontal increases when increasing the flow rate as shown in figure \ref{fig:tan1}. However, although this effect is important for narrow channels,  it almost disappears for the widest channel we have investigated.  For $W/d=570$, the surface slope remains almost constant close to the angle of repose of our material. This clearly shows that the increase of the slope with flow rate is due to the additional friction induced by the side walls as mentioned in several previous studies (\cite{taberlet03}; \cite{courrechdupont03}; \cite{gdrmidi04}).

\subsection{Free surface velocity}

\begin{figure}
  \begin{center}
  \includegraphics[scale=0.5]{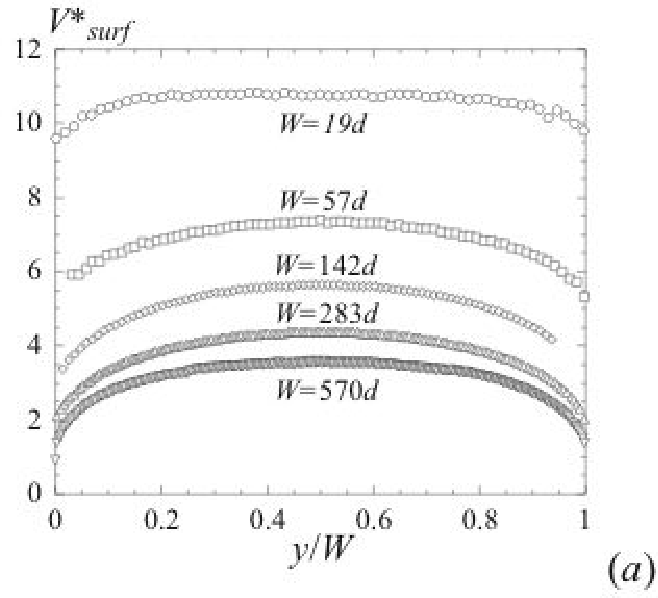}
  \includegraphics[scale=0.51]{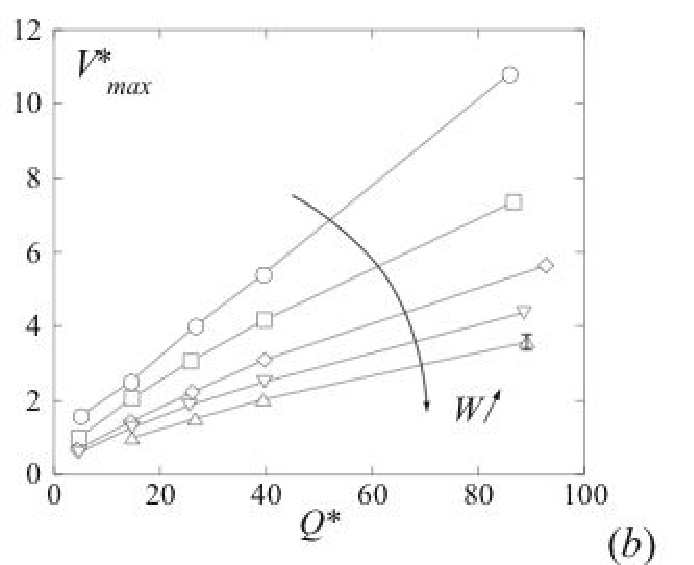}
  \includegraphics[scale=0.5]{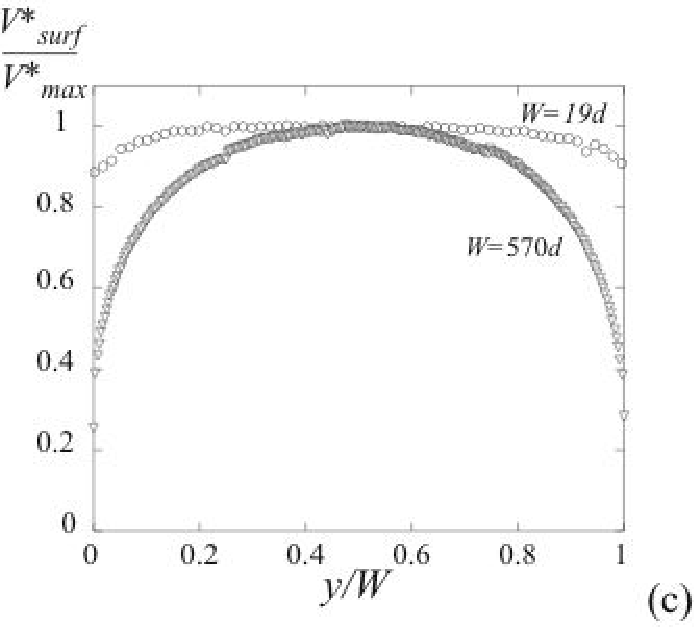}
  \includegraphics[scale=0.5]{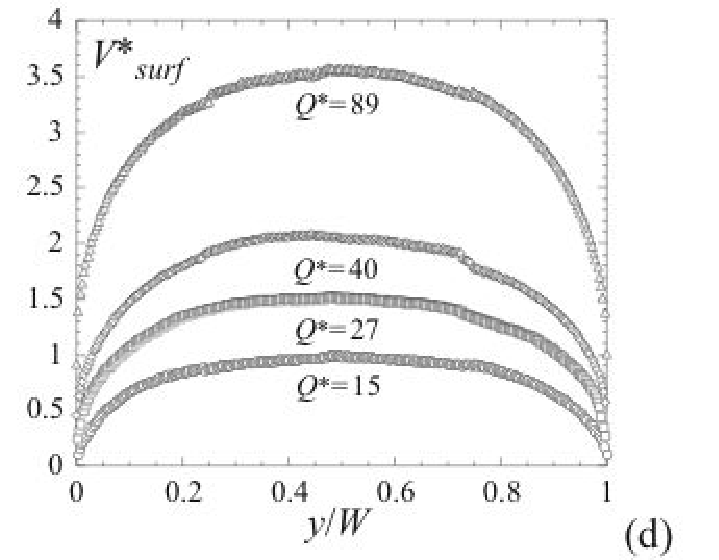}
  \caption{\label{fig:vel}  (\textit{a}) Dimensionless surface velocity profile $V^*_{surf}(y)$ as a function of $y/W$; for different widths $W^*$ at a given flow rate $Q^*=89$. (\textit{b}) Maximum surface velocity $V^*_{max}=V^*_{surf}(\frac{W}{2})$ as a function of $Q^*$. Same symbols as in figure \ref{fig:tan1}. (\textit{c})  Normalized velocity profiles $V^*_{surf}/V^*_{max}$ for $W^*=19$ and $570$. (\textit{d})  $V_{surf}^*(y)$ versus $y/W$ for different $Q^*$ for the widest channel $W^*=570$.}
  \end{center}
\end{figure}

In order to investigate the influence of the channel width on the free surface velocity, we first perform experiments keeping the same flow rate per unit of width $Q$  but changing the gap $W$ between the side walls.
 Figure \ref{fig:vel}(\textit{a}) shows  the  transverse surface velocity profiles  $V^*_{surf}(y)$ for $Q^*=89$.  In these figures the transverse direction $y$  is rescaled by the channel width.  The striking result is that the wider the channel, the slower the flow. By increasing the width from $19d$ to $570d$, the velocity is divided by a factor three. This effect is observed for all the flow rates as  shown in figure \ref{fig:vel}(\textit{b}). In this figure, the velocity $V^*_{max}$ in the centre line ($y/W=0.5$) is plotted as a function of the flow rate $Q^*$ for the different widths.  The velocity increases with respect to the flow rate but decreases when increasing the width. 
The last important remark about the surface velocity profile is that profiles are less and less uniform in the transverse direction when enlarging the channel. Whereas for $W^*=19$ a plug region with a constant velocity exists far from the walls,  this is no longer the case for $W^*=570$ (figure \ref{fig:vel}\textit{c}). In this widest channel, the profile is curved all across the channel whatever the flow rate is as shown in figure \ref{fig:vel}(\textit{d}). Therefore enlarging the channel does not allow to get rid of wall effects, contrary to what one would expect.

\begin{figure}
  \begin{center}
  \includegraphics[scale=0.49]{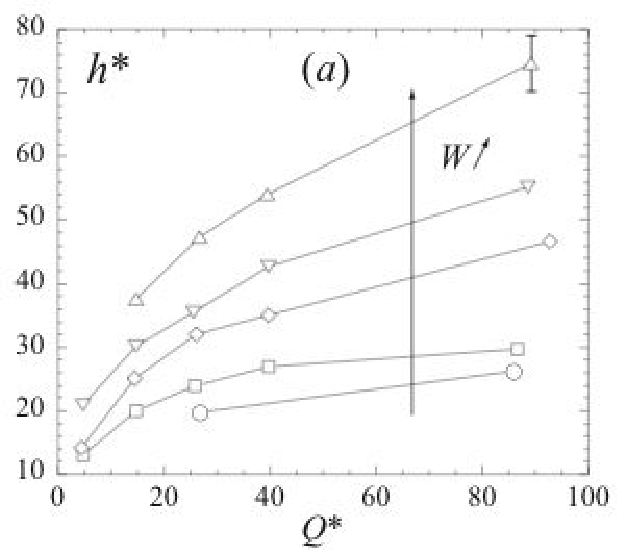}
  \includegraphics[scale=0.49]{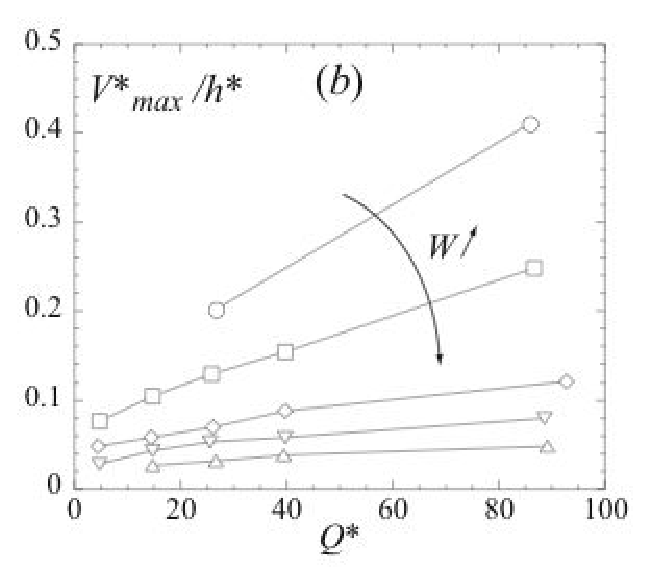}
   \caption{\label{fig:hgam}  (\textit{a}) Flow thickness $h^*$ measured in the centre line of the channel as a function of $Q^*$ for different widths. Same symbols as in figure \ref{fig:tan1}. (\textit{b}) Estimated shear rate $V^*_{max}/h^*$ as a function of $Q^*$ for different widths. Same symbols as in figure \ref{fig:tan1}.}
   \end{center}
 \end{figure}

\subsection{Thickness of the flow} 

We have seen above that, for a given flow rate per unit of width, the flow velocity decreases when enlarging the channel. Because of  mass conservation this implies  that the flow should be thicker. This is verified by our measurements of the thickness in the middle of the channel ($y/W=0.5$) using the erosion method. In figure \ref{fig:hgam}(\textit{a}), we have plotted how the thickness $h$ varies as a function of the flow rate for different channel widths $W$. The flow thickness increases with the flow rate but as expected from the velocity measurements, it also increases with $W$. Surprisingly,  for the widest channel the flow thickness reaches unusual high values about 75 grain diameters, much higher than the classical 10 or 20 particle diameters usually reported in the literature (\cite{gdrmidi04}). Concerning the variations of $h$ along $y$, no systematic measurement of the profile has been made. However, we have noticed a similar tendency than for the free surface velocity. For narrow channels, the flowing thickness is uniform across the width, whereas for wide channels, variations are observed: $h$ is smaller close to the wall than in the center part. For example for $W=142d$ and $Q^*=16.5$, $h=26.5d$ in the centreline of the channel but is only $17.5d$ at the wall.

\subsection{Discussion}

Whereas for narrow channels we recover results presented in the literature, \emph{e.g.}\,\,increase of inclination with flow rate, flowing layer about ten particle diameters thick, our measurements in wide channels dramatically contrast with the previous studies. The major difference is that the flow occurs on much thicker layer and at a much lower velocity in wide channels than in narrow channels. This is in contradiction with the result commonly admitted  for free surface flows that the shear rate in the flowing region of order of $0.6\sqrt{g/d}$ and weakly depends on the flow rate (\cite{khakhar01}; \cite{bonamy02a}; \cite{rajchenbach03}; \cite{gdrmidi04}). This is clearly shown in figure \ref{fig:hgam}(\textit{b}) where we have plotted an estimate of the shear rate $V^*_{max}/h^*$ as a function of the flow rate $Q^*$. Although for narrow channels the shear rate is of order $0.3\sqrt{g/d}$, it is almost 10 times lower in the widest channel. Moreover, figure \ref{fig:hgam}(\textit{b}) shows that the shear rate is not constant and obviously depends both on the flow rate $Q$ and on the channel width $W$.

This systematic study in channels of different widths therefore shows that side walls play a crucial role in the dynamics of granular flows on a pile. It seems impossible to get rid of them by enlarging the system. This observation suggests that it is essential to take into account side walls if one wants to model granular surface flows, at least in the steady regime. In the following we derive a theoretical model based on an empirical rheology recently proposed to describe dense granular flows, which takes into account additional wall friction, and compare the predictions with our experimental results.

\section{\label{theory}Theoretical model}

\subsection{Choice of the rheology}

In order to describe our surface flows, we need first to know the material rheology. Although the question of constitutive equations for dense granular flows is still a matter of debate,  recent works in different configurations seem to converge and allow to propose a simple empirical  rheology (\cite{gdrmidi04}). 

The first information comes from the inclined plane configuration, when a granular layer flows down an inclined bumpy surface. The study of the steady uniform flows for which scaling laws have been observed (\cite{pouliquen99a}; \cite{silbert03}),  allows to propose an empirical friction law to describe the shear stress that develops at the interface between the flowing layer and the rough surface. This law stipulates that the shear stress is proportional to the weight of the layer times a friction coefficient which depends on both the layer thickness $h$ and the mean depth-averaged velocity $\langle V \rangle$.  Based on the experimental measurements, the following form has been proposed for the basal friction coefficient $\mu_b(\langle V\rangle,h)$ (\cite{pouliquen02b}):

\begin{equation}
\mu_{b}(\langle V \rangle,h)      =    \mu_s  +     \frac{   \mu_2  -  \mu_s  }{ \displaystyle \frac{h \sqrt{gh}\beta}{\langle V \rangle  dL_0} + 1  },
\label{eq:mustop2}
\end{equation}
where $d$ is the particle diameter, $g$ the gravity and  $\mu_s$, $\mu_2$, $\beta$ and  $L_0$ are constants that depend
 on the material. This law has been successfully applied in the framework of depth-averaged equations to quantitatively predict the spreading of a granular mass (\cite{pouliquen02b}) and the development of instabilities (\cite{forterre03}). However, it cannot be considered as a constitutive law since it only applies at the base of the layer. 
 
 The link with the rheology has been made only recently (\cite{gdrmidi04}), when comparing these results with numerical studies in the configuration of the plane shear cell (Da Cruz {\it et al.} 2004; Iordanoff \& Khonsari 2004). In this configuration, a granular layer is confined between two rough plates under a confining pressure $P$ and sheared at a shear rate $\dot \gamma$. The numerical simulations reveal that the shear stress $\tau$ verifies a friction law and that the friction coefficient depends on a single dimensionless parameter $I$. $I$ can be interpreted as the ratio between the time scale given by the shear rate and the time scale related to the confining pressure (\cite{gdrmidi04}). The relation between shear stress and shear rate is then written in the following form:
\begin{equation}
\left |\frac{\tau}{P}\right |=\mu(I)   { \rm  \hspace{7 mm} with \hspace{7 mm}} I=\frac{|\dot \gamma | d}{\sqrt{P/\rho_s}},
\label{eq:mudeI}
\end{equation}
where $d$ is the particle diameter and $\rho_s$ is the particle density. 

 The interesting result arises when comparing this relation with the one obtained on the inclined plane for the basal friction law (\ref{eq:mustop2}). If one assumes that the material is everywhere defined by the local constitutive law given by relation (\ref {eq:mudeI}), predictions can be made for flows on inclined planes (see appendix \ref{app:rheolog}). One can then show that the predictions are compatible with the basal friction law (\ref{eq:mustop2}) issued from experimental measurements only if one chooses for the function $\mu(I)$ the following form:

\begin{equation}
\mu(I)=\mu_s+\frac{\mu_2-\mu_s}{I_0/I+1} \hspace{5mm}{\rm with}\hspace{5mm} I=\frac{|\dot \gamma|  d}{\sqrt{P/\rho_s}}.
\label{eq:muI}
\end{equation}

\begin{figure}
  \begin{center}
  \includegraphics[scale=0.7]{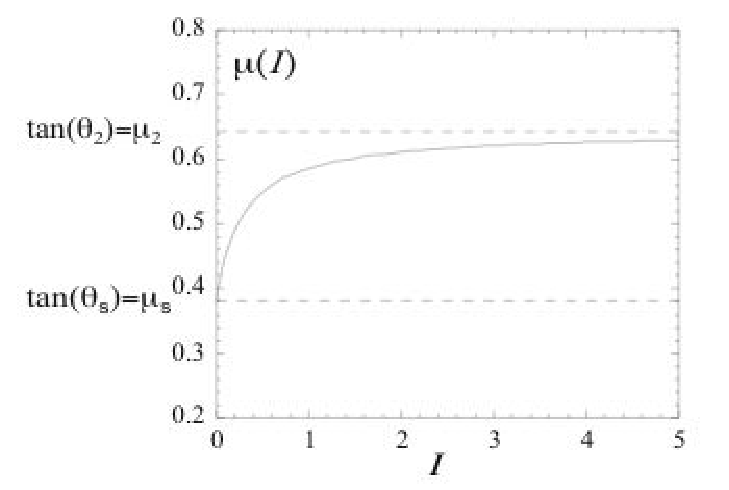}
  \caption{\label{fig:mu(I)} Friction coefficient $\mu$ as a function of the dimensionless parameter $I$ ($\mu_s=\tan(20.9^\circ)$, $ \mu_2=\tan(32.76^\circ)$, $ I_0=0.279$).}
 \end{center}
 \end{figure}

The coefficients $\mu_s$ and $\mu_2$ are the same as in relation (\ref{eq:mustop2}) and the constant $I_0$ is related to the coefficient $L_0$ and $\beta$ in (\ref{eq:mustop2}) (see appendix \ref{app:rheolog}).  
According to this law the friction coefficient goes from a minimum value $\mu_s$ for very low $I$ up to an asymptotical value $\mu_2$ when $I$ increases as sketched in figure \ref{fig:mu(I)}.

By interpreting the basal friction law found in inclined plane experiments in the framework of the constitutive law found in plane shear, we are then able to propose a simple local rheology.
The next step is to ask whether this rheology, which correctly describes plane shear and flows on inclined planes, can also predict surface flows on heaps. In the following, we apply relations (\ref{eq:mudeI}) and (\ref{eq:muI}) to heap flows taking into account the friction with side walls, and we compare the predictions with the experimental results presented in the previous section.  
In order to do so, we have to quantitatively determine the coefficients of the constitutive law (\ref{eq:muI}). 
The glass beads used in our study being the same as the ones used by Forterre and Pouliquen (2003) in  an inclined plane experiment, we can easily compute the coefficients of the relation $\mu(I)$ from the coefficients that have been measured for the basal friction law.  We found that $\mu_s=\tan(20.9^\circ)$,  $ \mu_2=\tan(32.76^\circ)$ and $ I_0=0.279$ (see appendix \ref{app:rheolog}).
This choice implies that there is no fit parameter in our constitutive law.  In other words, the idea adopted here is to calibrate the constitutive law on previous experiments on inclined plane, and check if quantitative predictions can be made for surface flows on heaps. 

\subsection{\label{sec:model}Model including wall effects}

\begin{figure}
 \begin{center}
  \includegraphics[scale=0.7]{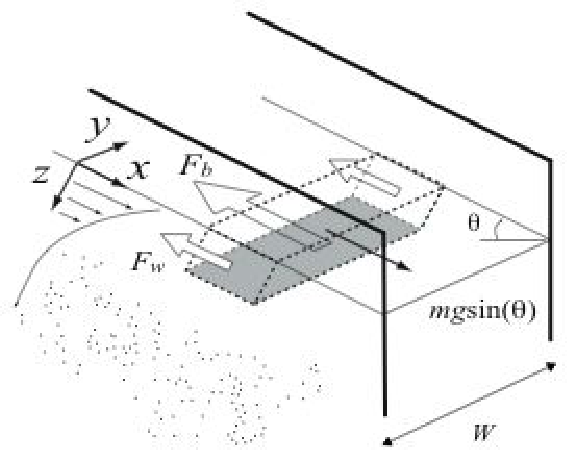}
  \caption{\label{fig:model} Force balance on an elementary slice of material.}
  \end{center}
 \end{figure}

Knowing the rheology, it is then possible to write the momentum balance for a granular layer  flowing on a heap. Let us consider a semi-infinite granular medium sandwiched between two smooth plates. We assume that the flow is uniform in the $x$ direction, the free surface making an angle $\theta$ with horizontal. In addition we neglect the variations of the velocity and the thickness in the $y$ direction. This simplification will be discussed in section \ref{limit}. We can then write the force balance for a slice of material of length $dx$, thickness $z$ and width $W$ (figure \ref{fig:model}).  
Four forces apply on the elementary slice: 
\begin{itemize}
\item the gravity: $dxW\rho gz\sin(\theta)$, where $\rho=\rho_s \phi$.
\item the two lateral friction forces due to the side walls $F_w=-2  dx \int_0^z \mu_w \rho gz' \cos(\theta) dz'$. Here two assumptions are made: first we assume that the beads slip against the side walls and we write the induce stress as a pure solid friction with a constant coefficient of friction $\mu_w$. Measurement of glass beads sliding on glass walls gives $\mu_w=\tan(10.4\pm 0.3^\circ)$. Second, the  pressure $P$ is assumed to be isotropic. This last assumption seems to be verified in simulations (\cite{prochnow00}; \cite{silbert01}).
\item the force $F_b$ that develops on the bottom face of the element due to the shear inside the material. According to our choice of the rheology (\ref{eq:mudeI}), this force is  $F_b=-dxW \mu(I(z))\rho g  z\cos(\theta)$, where $I(z)=|\dot \gamma (z)| /\sqrt{\phi gz\cos \theta}$.
\end{itemize}

The balance of the 3 forces leads to the following equation:
\begin{equation}
0=\tan(\theta) -\mu_w\frac{z}{W} -\mu(I(z)).
\label{eq:tan}
\end{equation} 

For a given inclination $\theta$, the parameter $I(z)$ and thus the shear rate $\dot\gamma(z)$ are then obtained from equations (\ref{eq:muI}) and (\ref{eq:tan}) and then integrated to give the velocity profile.  Before computing the velocity profile, a first remark can be made about the selection of the flow thickness. When going deeper in the pile -- increasing $z$ --, the friction term due to the walls -- the second term in (\ref{eq:tan}) -- increases. Consequently the force balance implies that $\mu(I(z))$ decreases. However,  the internal friction cannot be less than the critical value $\mu_s$ reached when $I$ goes to zero, \emph{i.e.}\,\,when the material is not sheared (figure \ref{fig:mu(I)}). As a consequence, there exists a critical depth $h$ below which the gravity, being screened by the lateral friction, is too weak to induce a shear in the material. The flow thickness is then given by the following relation:

\begin{equation}
\frac{h}{W}=\frac{\tan(\theta)-\mu_s}{\mu_w}.
\label{eq:htheta}
\end{equation}

This linear relation between the flow thickness and the channel width has been previously obtained by other authors (\cite{roberts69}; \cite{savage79}; \cite{taberlet03}). However, in our description where a local rheology is introduced,  we can go further and analytically derive from (\ref{eq:muI}) and (\ref{eq:tan}) all the flow properties: the velocity profile and the variations with the flow rate of the inclination, of the surface velocity and of the thickness. The exact analytical expressions are given in appendix \ref{app:analytic}. Examples of predicted velocity profiles are shown in figure \ref{fig:vexempl}. The flow is localized close to the free surface with a profile that looks like the ones experimentally observed.

Beyond the analytical results given in appendix \ref{app:analytic}, interesting scaling laws arise from the expression of the dimensionless parameter $I$. It is easy to show from equation (\ref{eq:tan}) that the shear rate can be written as follows: 

\begin{figure}
  \begin{center}
    \includegraphics[scale=0.7]{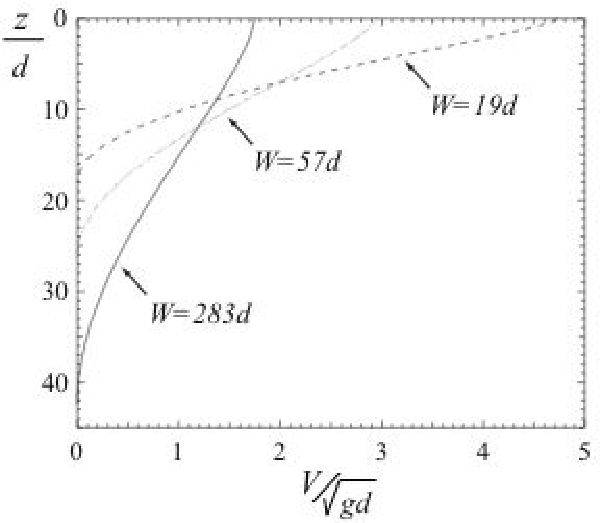}
    \caption{\label{fig:vexempl} Predicted velocity profiles along $z$ for a given flow rate and for different widths: $Q^*=31.5$ for $W=19d$ (dashed line), $W=57d$ (dotted line) and $W=283d$ (solid line).}
  \end{center}
\end{figure}

\begin{equation}
\frac{\dot{\gamma^*}(z^*,\theta)}{W^{*\frac{1}{2}}}=\sqrt{\frac{z^*}{W^*}}F\left (\frac{z^*}{W^*},\theta \right ),
\label{eq:gamma}
\end{equation}
where the function $F$ is given in the appendix \ref{app:analytic}.
By successive integrations we then get the following scaling laws for the velocity $V^*$ and flow rate $Q^*$ :
\begin{equation}
\frac{V^*(z^*,\theta)}{W^{*\frac{3}{2}}}=G\left (\frac{z^*}{W^*},\theta \right ) \hspace{5mm} {\rm and} \hspace{5mm}
\frac{Q^*(\theta)}{W^{*\frac{5}{2}}}=K\left (\theta \right ),
\label{eq:flowrate}
\end{equation}
where
\begin{equation}
G(x,\theta)=\int\limits^x_{\frac{h^*}{W^*}}\sqrt{Z}F(Z,\theta)dZ \hspace{5mm} {\rm and} \hspace{5mm} K(\theta)=\int\limits^{\frac{h^*}{W^*}}_0G(Z,\theta)dZ.
\label{eq:integrales}
\end{equation}
In all these expressions, the inclination $\theta$ plays the role of the control parameter. Since in our experiments we control the flow rate and not the inclination, it is convenient to rewrite the equations (\ref{eq:htheta}) and (\ref{eq:flowrate}) choosing $Q$ as a control parameter instead of $\theta$. The scaling laws then write as follow:
\begin{equation}
\tan(\theta)=f_1\left (\frac{Q^*}{W^{*\frac{5}{2}}}\right ),
\hspace{4mm}
\frac{V^*(z^*,Q^*)}{W^{*\frac{3}{2}}}=f_2\left (\frac{z^*}{W^*},\frac{Q^*}{W^{*\frac{5}{2}}} \right ) \hspace{4mm} {\rm and} \hspace{4mm}
\frac{h^*}{W^*}=f_3\left ( \frac{Q^*}{W^{*\frac{5}{2}}} \right ),
\label{eq:scaling}
\end{equation}
where the functions $f_1$, $f_2$ and $f_3$ can be linked to the functions $F$, $G$ and $K$ (see appendix \ref{app:analytic}). In terms of dimensional variables, the scaling laws are given by:
\begin{equation}
\tan(\theta)=f_1\left (\frac{dQ}{\sqrt{g}W^{\frac{5}{2}}}\right ),
\hspace{4mm}
\frac{dV(z,Q)}{\sqrt{g}W^{\frac{3}{2}}}=f_2\left (\frac{z}{W},\frac{dQ}{\sqrt{g}W^{\frac{5}{2}}} \right ) \hspace{4mm} {\rm and} \hspace{4mm}
\frac{h}{W}=f_3\left ( \frac{dQ}{\sqrt{g}W^{\frac{5}{2}}} \right ),
\label{eq:scalingdim}
\end{equation}

Let us underline that these scaling laws do not depend on the real shape of $\mu(I)$ (\ref{eq:muI}) but derive from the expression of the dimensionless number $I$. Only the functions $f_i$ are related to the exact expression of $\mu(I)$.

Finally, one can show from the forces balance (\ref{eq:tan}) and the shape of the friction coefficient $\mu(I)$ (\ref{eq:muI}) that steady uniform solutions are only possible for a finite range of surface inclinations given by $\mu_s<\tan\theta<\mu_2$. The lower limit corresponds to $h=0$, \emph{i.e.}\,\,the flow rate vanishes when $\theta\to\theta_s=\arctan(\mu_s)$. The upper limit corresponds to the maximal angle $\theta_2=\arctan(\mu_2)$ that can be balanced by the internal friction. Above this flow rate the flow accelerates along the pile. From appendix \ref{app:analytic} (\ref{eq:formulq2}), one can show that this maximal angle corresponds to a maximum flow rate $Q^*(\theta_2)$ for which steady flows are predicted:
\begin{equation}
\frac{Q^*(\theta_2)}{W^{*\frac{5}{2}}}=\frac{4}{15}I_0\sqrt{\phi \cos(\theta_2)}  \left (\frac{\mu_2-\mu_s}{\mu_w}\right)^{\frac{5}{2}}.
\label{eq:Qmax}
\end{equation}

\section{\label{comparison}Comparison with experimental measurements}
\subsection{Scaling laws}
\begin{figure}
  \begin{center}
 \includegraphics[scale=0.65]{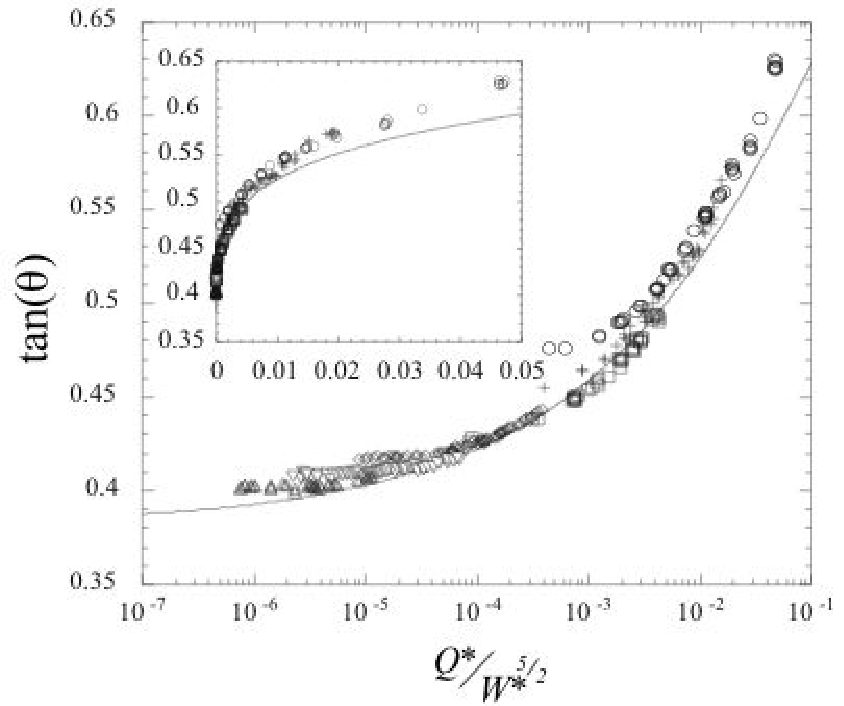}
  \caption{\label{fig:scalingtan}  $\tan{\theta}$ as a function of $Q^*/W^{*5/2}$ for different widths ($W^*$ from 19 to 570), same data as in figure \ref{fig:tan1}. Inset: same plot in lin-lin axis. The solid line is the prediction of the model (\ref{eq:scaling}).}
  \end{center}
 \end{figure}

\begin{figure}
  \begin{center}
 \includegraphics[scale=0.650]{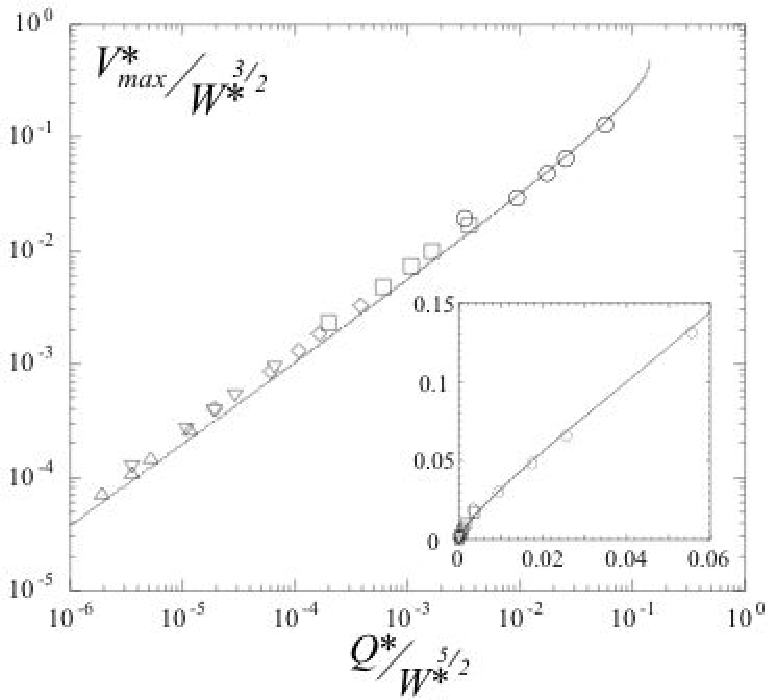}
  \caption{\label{fig:scalingV}  The rescaled maximum surface velocity $V^*_{max}/W^{*3/2}$ as a function of $Q^*/W^{*5/2}$ for different widths. Inset: same plot in lin-lin axis. The solid line is the prediction of the model (\ref{eq:scaling}).}
  \end{center}
 \end{figure}

\begin{figure}
  \begin{center}
 \includegraphics[scale=0.650]{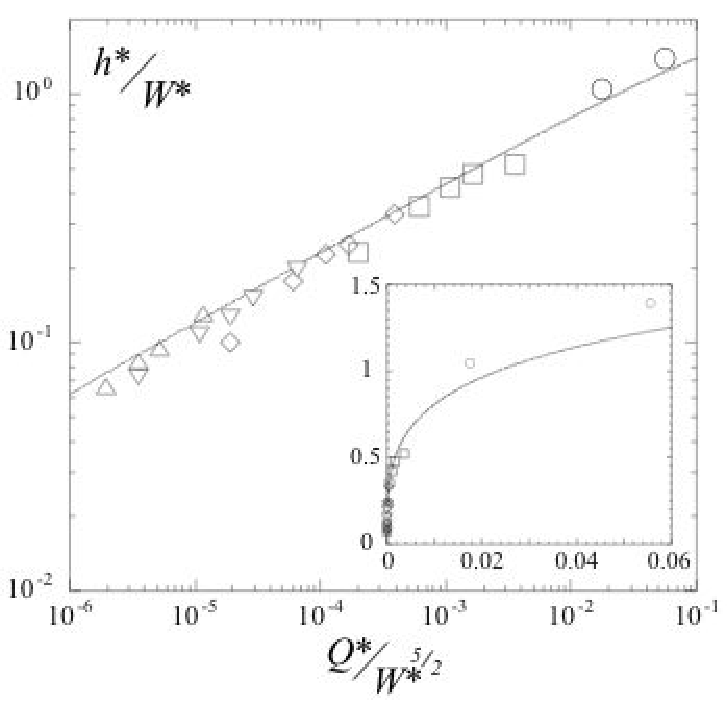}
  \caption{\label{fig:scalingh}  Rescaled flowing thickness $h^*/W^*$ as a function of $Q^*/W^{*5/2}$. Inset: same plot in lin-lin axis. The solid line is the prediction of the model (\ref{eq:scaling}).}
  \end{center}
 \end{figure}

In order to check the validity of the theoretical model, we first compare all our experimental measurements carrying out for different widths and different flow rates with the predicted scaling laws (\ref{eq:scaling}). To this end, we have plotted in figures \ref{fig:scalingtan}, \ref{fig:scalingV} and \ref{fig:scalingh} our measurements in terms of the three quantities $\tan(\theta)$, $V_{surf}^*/W^{*\frac{3}{2}}$ and $h^*/W^*$ as a function of $Q^*/W^{*\frac{5}{2}}$, the scaling predicted by relations (\ref{eq:scaling}). The striking result is that, in the three cases, all the data obtained from different channel widths collapse on a single curve, showing that the predicted scalings correctly capture the influence of both the flow rate and the channel width. The collapse is perhaps less accurate in figure \ref{fig:scalingtan} for the inclination, where each set of  data obtained for different $W$ seems to converge  to a different angle for $Q^*=0$.

The second main result is that, not only collapse the experimental data, but they also quantitatively follow the theoretical predictions, as shown in figures \ref{fig:scalingtan}, \ref{fig:scalingV} and \ref{fig:scalingh}, where the solid lines correspond to predictions. The agreement between the model and the experimental data is quantitatively good, although one can notice that the predictions for the surface velocity are  15-20$\%$ below the measurements and those for the thickness $h$ are 10-15$\%$ above the measurements. Another quantitative test consists in studying the asymptotic behaviour at low flow rates. Power laws can be predicted by developing relations (\ref{eq:scaling}) for small flow rates as shown in the appendix \ref{powerlaw}. One finds the following relations:

\begin{equation}
\frac{V_{surf}^*}{W^{*\frac{3}{2}}} \approx \left (\frac{Q^*}{W^{*\frac{5}{2}}} \right )^{\frac{5}{7}} \hspace{3mm} {\rm and} \hspace{5mm}
\frac{h^*}{W^*}\approx \left ( \frac{Q^*}{W^{*\frac{5}{2}}} \right )^{\frac{2}{7}}
\hspace{5mm} {\rm when} \hspace{5mm}
\left ( \frac{Q^*}{W^{*\frac{5}{2}}} \ll 1\right ).
\label{eq:powerlaw}
\end{equation}
When comparing with the experimental measurements the agreement is good. The fit of our data gives  for the surface velocity an exponent equal to $0.722$ to be compared with $5/7\approx 0.714$ and  for the thickness an exponent equal to $0.295$ to be compared to $2/7\approx 0.286$.

\begin{figure}
  \begin{center}
 \includegraphics[scale=0.70]{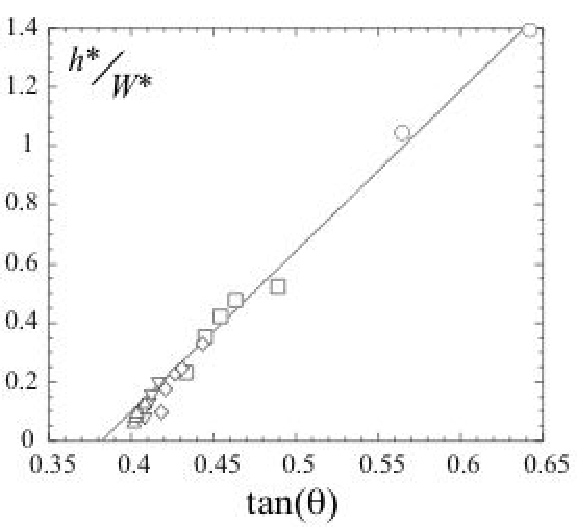}
\caption{\label{fig:scalinghtan}  Rescaled thickness $h^*/W^*$ as a function of $\tan(\theta)$ for different widths $W^*=$ 19 ($\circ$), 57 ($\Box$), 142 ($\Diamond$), 283 ($\bigtriangledown$), 570 ($\triangle$). The solid line is the prediction of the model (\ref{eq:htheta}).}
  \end{center}
 \end{figure}

The last prediction that can be checked is the relation between the thickness $h/W$ and the angle $\theta$ (\ref{eq:htheta}) coming from the force balance. The model predicts a linear variation of the thickness with the tangent of the inclination. Figure \ref{fig:scalinghtan} shows that our measurements are in good agreement with the theoretical results. The experimental data obtained for different widths collapse on a single straight line, with a slight departure from linearity observed at low flow rates. A good quantitative agreement is again obtained.

This analysis then shows that the characteristics of steady uniform flows on a heap are well predicted by our model based on the rate dependent friction law coupled with wall effects. It is important to keep in mind that in this analysis no parameter is fitted. The parameters of the rheological law $\mu(I)$ have been chosen from the previous study by Forterre \& Pouliquen (2003) who used the same granular material. The same rheology is therefore able to quantitatively describe flows on a rough inclined plane and surface flows on a heap.
 
 In the next section we try to go further and discuss the domain of existence of steady uniform flows, by comparing the range of flow rates where they are observed with the prediction of the theory.
  
\subsection{\label{thresholdmodel}Minimum flow rate}

In section 4.2 we have seen that the model predicts that steady uniform flows are possible for a range of flow rates going from zero, when the thickness vanishes and $\theta\to\theta_s$, to a maximum value when $\theta\to\theta_2$. We have not been able in our experiment to reach high enough flow rates to conclude about the existence or not of a maximum flow rate. However, our measurements clearly demonstrate the existence of a minimum flow rate for steady uniform flows, below which the flow is intermittent. This minimum flow rate depends on the channel width as shown in figure \ref{fig:qc}. 

This transition between steady flows and intermittent flows is not predicted by the model. However, it is interesting to compare this situation with results about flow threshold obtained in the inclined plane configuration. It has been shown that, when a granular layer flows down a rough plane inclined at an angle $\theta$ without any side wall, no flow is possible if the layer thickness is less than a critical thickness $h_{stop}(\theta)$ (\cite{pouliquen99a}; \cite{daerr01}; \cite{silbert03}). 
The existence of a critical thickness seems to be linked to the existence of correlations in the grain motions (\cite{ertas02}; \cite{gdrmidi04}; \cite{pouliquen04}). Explaining the  flow threshold of granular material is far beyond the scope of this study. However, it is interesting to wonder if a connection exists between the minimum thickness observed in the inclined plane configuration and the minimum flow rate we observe for flows on  heap. When decreasing the flow rate on a heap, the flow thickness decreases. The critical flow rate can then be interpreted as a critical thickness. How does it compare with the minimum thickness $h_{stop}(\theta)$ observed on inclined plane?

\begin{figure}
  \begin{center}
 \includegraphics[scale=0.650]{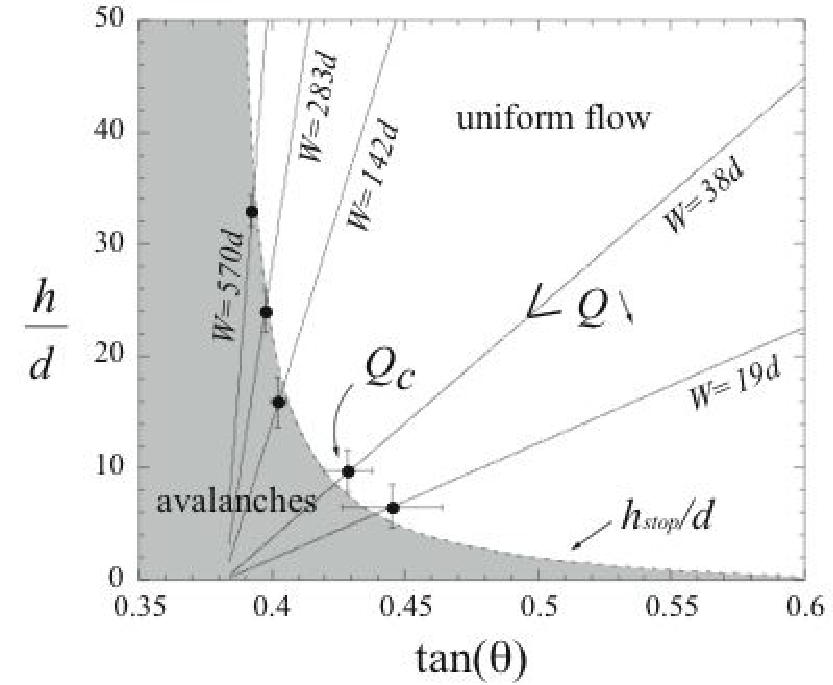}
  \caption{\label{fig:scalinghc} Comparison of the domain of steady uniform flows between flows on a heap (black dots, same data as in figure \ref{fig:qc}) and flows on inclined plane (dashed line, function $h_{stop}/d$ from Forterre \& Pouliquen (2003) ). The continuous straight lines indicate how $h/d$ and $\tan\theta$ vary in the heap flow configuration when varying the flow rate $Q$ at a fixed channel width $W$.}
  \end{center}
 \end{figure}

In order to answer this question, we consider a flow on a pile  in a channel of width $W$.  For a given flow rate $Q$, the inclination of the free surface $\theta$  and the thickness $h$ are  given in our model by equations (\ref{eq:scaling}). They are related by the linear relation (\ref{eq:htheta}) such that in the  plane $(\tan\theta, h/d)$ the system follows a straight line when varying $Q$. In figure \ref{fig:scalinghc}, five of such lines are drawn corresponding to five different channel widths. In the model, each point of the lines corresponds \textit{a priori} to a flow. However, experimentally this is not the case. Because there exists a critical flow rate below which the flow is intermittent, only the upper part of the lines in figure \ref{fig:scalinghc} can be reached. Our measurement of the critical flow rate $Q_c(W)$ then gives a critical thickness $h_c(W)$ and a critical inclination $\theta_c(W)$, which are plotted in figure \ref{fig:scalinghc} for five widths.
The striking results comes from the comparison between the positions of the points $(\tan\theta_c(W),h_c(W))$ and  the flow threshold $h_{stop}(\theta)$ observed on inclined plane. In figure \ref{fig:scalinghc}, we have plotted as a dashed line the frontier between flow and no flow  $h_{stop}(\theta)/d$ coming from the inclined plane experiments by Forterre \& Pouliquen (2003) using the same glass beads. One observes that the transition points  $(\theta_c(W),h_c(W))$ between steady and intermittent flows on a heap coincide with the frontier $h_{stop}(\theta)$ between flow and no flow in inclined plane.
This result  shows that the two \textit{a priori} different conditions to get  steady uniform flows in both configurations, heap flow and inclined plane flows, are  {\it quantitatively} the same.
 
\section{\label{discussion}Discussion}
\subsection{\label{limit}Limits of the model}

The theoretical model, which takes into account both the specificity of the granular rheology and the influence of lateral walls, gives good quantitative predictions for the steady uniform flows observed on heap. However, although the major features are captured, some discrepancies exist which give information about the limits of the approach. 

The first limit is that we have written the forces balance assuming the flow to be uniform across the channel. This choice, which simplifies the theoretical description, does not allow to describe the velocity variations experimentally observed across the channel. If one wants to capture these three-dimensional effects, one has to take into account shear in the transverse direction. It is then necessary to write a  generalization of the constitutive law (\ref{eq:mudeI}) for fully three-dimensional deformations.

A second limit concerns the use of the local rheology written as a simple rate dependent friction law. As discussed in GDR MiDi (2004), this local approach is probably no longer valid close to the flow threshold or close to the boundaries. This limit is indeed observed in our case and gives rise to several discrepancies between the prediction and the measurements. The first one concerns the flow threshold. The local rheology is not able to predict the critical flow rate observed experimentally below which intermittent flows occur. We have been able to understand the existence of this minimum flow rate only by stipulating \textit{a priori} that the flow cannot occur below a critical thickness, by analogy to the case of flows on inclined planes. The second discrepancy concerns the velocity profiles. The predicted profile presents a zero shear rate at the free surface and a static region with zero velocity deep inside the pile. Experimentally the velocity profile exhibits a non zero shear rate at the surface and an exponential tail deep in the pile (\cite{gdrmidi04}). One possible explanation for the limit of the local rheology is that close to the flow threshold or close to boundaries, non local phenomena have to be taken into account. Grain motions are correlated on lengths which become comparable to the flow thickness or to the distance to the boundaries (\cite{ertas02}; \cite{gdrmidi04}; \cite{pouliquen04}). The precise understanding of this limit and the development of constitutive laws taken into account the non-local effects are still a challenge. 

The last limit of our model lies in the description of the interaction of the flowing layer with the side walls. The stress that develops between the material and the glass wall is described in our model as a simple Coulomb friction. Whereas it is reasonable for slow flows, for fast flows collisions between particles and wall could  occur and  become predominant, leading to a rate dependent wall stress. This limitation could affect the maximum inclination predicted in our theory. According to our model, no steady uniform flow on a pile can occur above the angle $\theta_2$. This value is typically between $30$ and $40$ degrees depending on the material used (\cite{forterre03}). Recent experiments by Taberlet \textit{et al.} (2003) report much higher slopes up to 60 degrees in narrow channels and at high flow rates. A plausible explanation for such high inclinations could be that the stress at the wall becomes collisional. Our model should then be modified to take into account a rate dependent wall friction.

\subsection{Transition between flow on an inclined plane and flow on a pile}

\begin{figure}
  \begin{center}
 \includegraphics[scale=0.6]{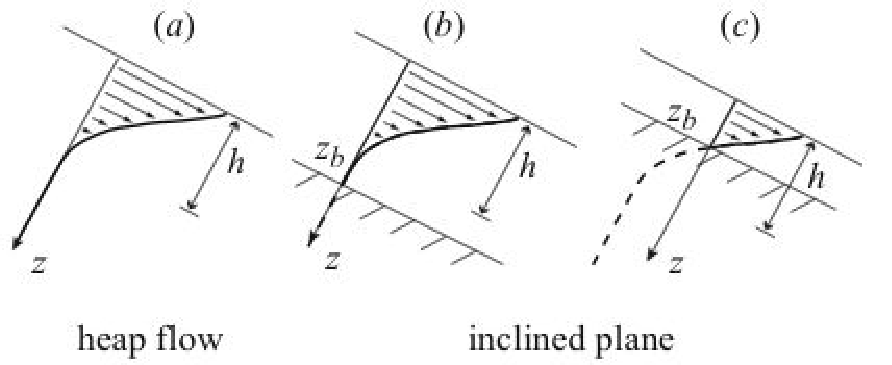}
  \caption{\label{fig:transition}  Modification of the velocity profile when a rigid bottom is introduced in the model for a given angle. (\textit{a}) No bottom. (\textit{b}) The rough bottom is deeper than the zero velocity level $z_b>h$. (\textit{c}) $z_b<h$. The dashed lines correspond to the velocity profile in the case of semi-infinite flow.}
  \end{center}
\end{figure}

  \begin{figure}
  \begin{center}
 \includegraphics[scale=0.7]{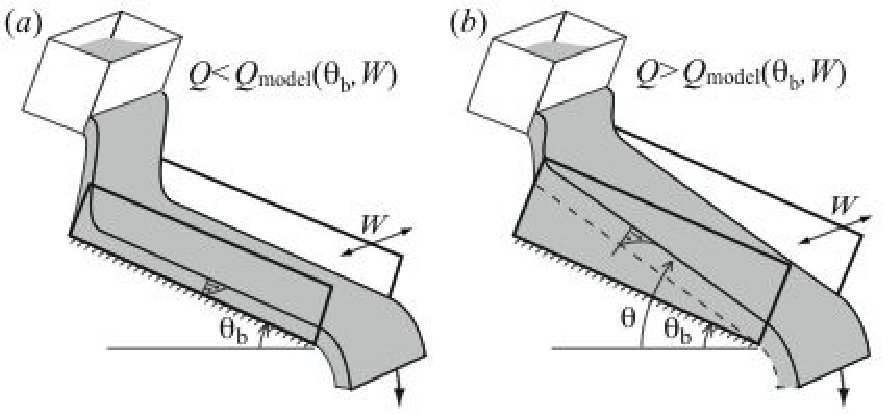}
  \caption{\label{fig:transition2} Sketch of the transition between (\textit{a}) flow on rough inclined plane and (\textit{b}) heap flow when increasing the flow rate.}
  \end{center}
\end{figure}
 
In the theory presented in section \ref{sec:model}, the medium is considered as semi-infinite in order to model flows on a pile. However, flows of a finite layer corresponding to inclined plane experiments carried out in narrow channels can be described using the same approach. One just has to stipulate that the velocity should vanished at a given depth $z_b$ in order to reproduce the no slip condition on a rough surface. If the bottom plate is located deeper than the critical thickness predicted in the semi-infinite case, the bottom does not introduce any modification. The flow is made of a static layer close to the rough surface with a flowing region on top of it (figure \ref{fig:transition}b). However, when the bottom plate is close to the free surface, the model predicts that the whole granular layer is flowing. The velocity profile is therefore given by the upper part of the velocity profile for an infinite heap (figure \ref{fig:transition}c). This simple observation allows to better understand  the formation of static piles observed in experiments on inclined planes in narrow channels. Savage (1979), Ancey (2001) and Taberlet \emph{et al.} (2003) have shown that when the inclination is low or the flow rate is high, the material does not flow down to the bottom plate but a static pile develops. A transition is then observed between flow on a plane and flow on a pile. This can be understood as follow. In an inclined plane experiment the two control parameters are  the inclination of the plane and the flow rate. For a given inclination, if the flow rate is less than the flow rate predicted in the case of semi infinite pile at this angle, we are in case of figure \ref{fig:transition}(\textit{c}). The flow extends over the whole granular layer, the free surface being parallel to the bottom plane (figure \ref{fig:transition2}\textit{a}).  However, when increasing the flow rate, the thickness increases and the friction with the wall becomes more and more important. Eventually, the only way for the system to overcome the side wall friction is to increase the free surface inclination at an angle higher than the bottom inclination $\theta_b$ (figure \ref{fig:transition2}\textit{b}). The formation of a static pile in confined inclined planes is thus controlled by the side walls. This explains why in very wide set-ups (\cite{pouliquen99a}), or in molecular dynamic simulations using periodic boundary conditions (\cite{chevoir01}; \cite{silbert03}), no static layer was ever observed, as soon as the inclination of the plate is above the repose angle of the material.

\section{\label{conclusion}Conclusion}

In this paper, steady uniform flows obtained when granular material is released at the top of a pile have been investigated. The influence of the side walls on these surface flows has been carefully studied, and a theoretical model has been proposed. From the experimental and theoretical analysis, two main conclusions emerge.

The first conclusion is that steady uniform flows on a pile are entirely controlled by side wall effects. The characteristics of flows in wide channels differ notably from the ones observed in narrow channels. For the same flow rate per unit of width,  the flow is thicker and slower in a wide channel than in a narrow channel,  thicknesses up to 70 particle diameters being observed. These observations contrast dramatically with the picture commonly accepted that the flowing zone is necessarily thin with a shear rate roughly constant. It means that precautions have to be taken when interpreting experimental results in narrow channels to get information about the intrinsic flow rheology.

The second result concerns the constitutive law for dense granular flows. Using the local rheology recently proposed in GDR MiDi (2004) and taking into account the friction due to the lateral walls, we have been able to reproduce the major characteristics of granular heap flows. In particular the localisation of the flow at the free surface is predicted, but appears to be a consequence of the wall friction and is not an intrinsic property of the rheology. Our analysis therefore dispels the difficulty raised in the article GDR Midi (2004), that flows on inclined planes and flows on a heap are not compatible when side wall effects are neglected. We have shown here that both configurations can actually be unified in the same framework choosing the rheology $\mu(I)$. The unification is not only qualitative but also quantitative. The flow thickness, the velocity and the free surface inclination measured for flow on a heap are predicted within 15 $\%$, when  the friction law is calibrated based on measurements for  flows on inclined planes. The comparison also extends to the flow threshold, which appears to be the same in both configurations once the critical flow rate observed in heap flows is interpreted in terms of a critical thickness.
 Of course, number of limits exists that has been identified and related to the fact that the proposed constitutive law is local. However, we believe that this approach represents a very serious candidate for a constitutive law for dense granular flows, as it is able to describe in the same framework three different configurations: Couette flows, inclined planes, and flows on a heap.

This study, which contrasts with previous results on surface flows, raises many questions. The first one concerns the generalisation of the constitutive law to three-dimensional deformations. We have seen that the approach proposed in this paper does not allow to describe the deformation observed in the transverse direction. What kind of constitutive law has to be written to capture this variation?

The second question concerns the role of the channel length. We have seen that in the steady uniform regime the flow thickness is controlled by the width of the channel. What would happen for very wide channels, when the width of the flow becomes comparable to the total length of the flow? Is the flow length in this case the new relevant length scale of the problem?

 Another question  concerns unsteady and non-uniform flows. What will happen for transient flows for example when avalanches are triggered on top of a pile? Do the side walls still play an important role in this case? To describe the complex dynamics of avalanches, several models have been proposed based on depth-averaged hydrodynamics equations (\cite{bouchaud94}; \cite{boutreux98}; \cite{douady99}; \cite{khakhar01b}; \cite{aradian02}).  What is the link between these approaches and the model presented here? Is it possible to get deeper insight on these models, now that we have more information about the internal rheology suitable for dense granular flows? These questions represent work for future investigations.
 
Finally, a last but fundamental issue concerns the physical origin of the constitutive law. Up to now the proposed rheology rests on an empirical ground. Attempts to link the observed macroscopic behaviour to the microscopic grain motion exist (\cite{mills99}; \cite{pouliquen01}; \cite{ertas02}; \cite{gdrmidi04}) which are based on the idea that grains experience correlated or cluster-like motions. Recent experiments seem to support this idea (\cite{pouliquen04}; \cite{choi04}), but the derivation of constitutive laws from microscopic bases remains an open challenge.

\begin{acknowledgments}

The authors thank Bruno Andreotti for enlightening discussions and advices for the PIV method and Fr\'ed\'eric Ratouchniak for technical support. 
\end{acknowledgments}

\appendix

\section{\label{app:rheolog}Rheological law}
In this appendix we show how the local constitutive law $\mu(I)$ (\ref{eq:muI}) can be obtained from the expression of the basal friction law obtained in experiments on inclined planes (\cite{pouliquen02b}; \cite{forterre03}).

To this end we consider a granular layer of thickness $h$ flowing on a plane inclined at an angle $\theta$. On one hand the experiments on steady uniform flows have shown that the bottom friction can be expressed in term of the thickness $h$ and the depth-averaged velocity $\langle V \rangle$ as follow:

\begin{equation}
\mu_{b}(\langle V \rangle,h)      =    \mu_s  +     \frac{   \mu_2  -  \mu_s  }{\displaystyle \frac{\beta h \sqrt{g h} }{\langle V \rangle L_0} + 1  },
\label{eq:mustop}
\end{equation}
where $\mu_s$,  $\mu_2$, $\beta$ and $L_0$ are constants.
One the other hand, if one assumes that the material is described by the constitutive law $\mu(I)$ with $I=|\dot\gamma| d/\sqrt{P/\rho_s}$, the force balance in the bulk implies $\mu(I(z))=\tan(\theta)$. This implies that the parameter $I$ is constant across the layer, independent of $z$,  depending only on the inclination of the plane. As a consequence, from the definition of $I$, we obtain the velocity profile (\cite{gdrmidi04}):

\begin{equation}
\frac{V(z)}{\sqrt{gd}}=\frac{2}{3}I(\theta)\sqrt{\phi \cos(\theta)}\frac{(h^{3/2}-z^{3/2})}{d^{3/2}}.
\label{eq:velocity3}
\end{equation}

From this equation we can compute the depth-averaged velocity: 

\begin{equation}
\frac{\langle V\rangle}{\sqrt{gh}}=\frac{2}{5} I(\theta) \sqrt{\phi \cos(\theta)}\frac{h}{d}
\label{eq:velocity4}.
\end{equation}

From this result we can then rewrite the basal friction coefficient (\ref{eq:mustop}) in terms of the parameter $I$ as follow:
\begin{equation}
\mu_b(I)   =     \mu_s  +     \frac{   \mu_2  -  \mu_s  }{  \displaystyle \frac{5\beta d }{2L_0 I\sqrt{\phi \cos \theta}} + 1  }.
\label{eq:formulmub}
\end{equation}

For consistency reason, the friction law $\mu(I)$ is then given by:
\begin{equation}
\mu(I)   =     \mu_s  +     \frac{   \mu_2  -  \mu_s  }{   \frac{I_0 }{I} + 1  }, 
\label{eq:muIapp}
\end{equation}
with
\begin{equation}
I_0    =  \frac{5}{2}\frac{d\beta}{L_0\sqrt{\phi \cos(\theta)}}.
\label{eq:formulI0}.
\end{equation}
 
In the constitutive law $I_0$ should be a constant, whereas it depends on $\theta$ through the term $\sqrt{ \cos(\theta)}$. However, this term does not vary much in the experimental range, such that one can think that it is missing in the expression (\ref{eq:mustop}) coming from the experimental data. 
Finally, from experimental measurements, Forterre and Pouliquen (2003) give $L_0/d=1.65$, $\beta=0.136$, $\mu_s=\tan(20.9^\circ)$, $\mu_2=\tan(32.76^\circ)$. Taking $\phi \approx 0.6$ and  an average value of $25^\circ$ for $\theta$, we obtain $I_0=0.279$. The parameters $\mu_s$, $\mu_2$ and  $I_0$ then quantitatively define the constitutive law. 

\section{\label{app:analytic}Analytical expressions for velocity profiles and for flow rate}

Here we derive the expressions of the velocity profile and the flow rate as a function of the angle $\theta$ and of the depth $z$.
The local friction law (\ref{eq:muI}) and the forces balance equation (\ref{eq:tan}) give the following equation for the shear rate:
\begin{equation}
\frac{\dot{\gamma^*}(z^*)}{W^{*\frac{1}{2}}}=-I_0\sqrt{\frac{z^*}{W^*}\phi\cos(\theta)}\left (\frac{\tan(\theta)-\mu_w\frac{z^*}{W^*}-\mu_s}{\mu_2-\tan(\theta)+\mu_w\frac{z^*}{W^*}}\right )\equiv \sqrt{\frac{z^*}{W^*}}F\left (\frac{z^*}{W^*},\theta\right ).
\label{eq:formulgam1}
\end{equation}
The minus sign before $I_0$ comes from the sign of the shear stress in our configuration. 
To simplify the expressions, we define:

\begin{equation}
\frac{h^*}{W^*}=\frac{\tan(\theta)-\mu_s}{\mu_w} \;{\rm , }\hspace{5mm} \frac{h_2^*}{W^*}=\frac{\mu_2-\tan(\theta)}{\mu_w}\;{\rm ,  }
\hspace{5mm} {\rm     and  }\hspace{5mm} 
\Delta\mu=\mu_2-\mu_s,
\label{eq:simplify}
\end{equation}
which leads to:
\begin{equation}
\frac{\dot{\gamma^*}(z^*)}{W^{*\frac{1}{2}}}=-I_0\sqrt{\frac{z^*}{W^*}\phi\cos(\theta)}\left (\frac{h^*-z^*}{h_2^*+z^*}\right ).
\label{eq:formulgamlight}
\end{equation}
Integrating the above equation with the boundary condition $V^*=0$ at $z^*=h^*$, we find:

\begin{eqnarray}
\label{eq:formulv3}
\frac{V^*(z^*,\theta)}{W^{*\frac{3}{2}}}  & = &2I_0  \sqrt{ \phi  \cos(\theta)} \\ \nonumber
& & \times
\bigg{ [ } \frac {\Delta \mu}{\mu_w}   \left ( \sqrt{\frac{h^*}{W^*}} - \sqrt{\frac{z^*}{W^*}} - \sqrt{\frac{h_2^*}{W^*}}\left ( \arctan\sqrt{\frac{h^*}{h_2^*}} - \arctan\sqrt{\frac{z^*}{h_2^*}}\right ) \right ) \\ \nonumber
 & & -  \frac{1}{3} \left ( \frac{h^*}{W^*} \right )^\frac{3}{2}  + \frac{1}{3} \left ( \frac{z^*}{W^*} \right )^\frac{3}{2}  \bigg{ ] }\\ \nonumber
 & \equiv & G \left (\frac{z^*}{W^*},\theta\right ) \\ \nonumber
 &\equiv&
 G_2\left (\frac{z^*}{W^*}, \frac{h^*(\theta)}{W^*}\right ).
\end{eqnarray}

Integrating the previous equation from the bottom of the flowing layer ($z^*=h^*$) to the free surface yields:

\begin{eqnarray}
\label{eq:formulq2}
\frac{Q^*(\theta)}{W^{*\frac{5}{2}}}  &= & \frac{V_{surf}^*(\theta)}{W^{*\frac{3}{2}}} \frac{h^*}{W^*} + 2I_0  \sqrt{ \phi  \cos(\theta) }\left ( \frac{h_2^*}{W^*}\right ) ^{\frac{3}{2}}  \times \\ \nonumber
& & \left \{ \frac{2}{15} \frac{h_2^*}{W^*}\left ( \frac{h^*}{h_2^*}\right )^\frac{5}{2} - \frac{2}{3} \frac{\Delta\mu}{\mu_w} \left ( \frac{h^*}{h_2^*}\right )^\frac{3}{2} + \frac{\Delta\mu}{\mu_w}\left [ 
\left ( \frac{h^*}{h_2^*}+1\right ) {\arctan \sqrt{\frac{h^*}{h_2^*}}} - \sqrt {\frac{h^*}{h_2^*}}
 \right ]   \right \} \\ \nonumber
 &\equiv& K\left (\theta\right )\\ \nonumber
 &\equiv&
 K_2\left ( \frac{h^*(\theta)}{W^*}\right ).
\end{eqnarray}
 
where $V_{surf}$ is the velocity at the free surface ($z^*=0$). Equations (\ref{eq:simplify}-left), (\ref{eq:formulv3}) and (\ref{eq:formulq2}) can be rewritten with the following form:
\begin{equation}
\tan(\theta)=f_1\left (\frac{Q^*}{W^{*\frac{5}{2}}}\right ),
\hspace{4mm}
\frac{V^*(z^*,Q^*)}{W^{*\frac{3}{2}}}=f_2\left (\frac{z^*}{W^*},\frac{Q^*}{W^{*\frac{5}{2}}} \right ) \hspace{4mm} {\rm and} \hspace{4mm}
\frac{h^*}{W^*}=f_3\left ( \frac{Q^*}{W^{*\frac{5}{2}}} \right ),
\label{eq:scalingapp}
\end{equation}
where:
\begin{equation}
f_1(X)=\tan \left ( K^{-1}\left ( X \right ) \right ), \hspace{2mm} f_2(X,Y)=G_2\left ( X , K_2^{-1} \left ( Y \right ) \right )\hspace{2mm} {\rm and} \hspace{2mm}
 f_3(X)=K_2^{-1}\left(X\right ).
\label{eq:f1}
\end{equation}

\section{\label{powerlaw}Asymptotic expressions at low flow rates}

In this appendix we give the asymptotic relations between (\ref{eq:Qmax}) for low flow rate, \emph{i.e.}\,\,$Q^*/W^{*5/2}\rightarrow 0$, $h^*/W^*\rightarrow 0$ and $\theta\rightarrow\theta_s$. Then (\ref{eq:formulv3}) and (\ref{eq:formulq2}) at the lower order give:

\begin{equation}
\frac{V_{surf}^*}{W^{*\frac{3}{2}}}  \approx \frac{4}{15}I_0  \sqrt{ \phi  \cos(\theta_s)} \frac{\Delta\mu}{\mu_w}\left (\frac{h^*}{W^*}\right )^{\frac{5}{2}},
\label{eq:approxvh}
\end{equation}

\begin{equation}
\frac{Q^*}{W^{*\frac{5}{2}}}  \approx \frac{4}{35}I_0  \sqrt{ \phi  \cos(\theta_s)} \frac{\mu_w}{\Delta\mu}\left (\frac{h^*}{W^*}\right )^{\frac{7}{2}},
\label{eq:approxqh}
\end{equation}

which can be rewritten as follow:

\begin{equation}
\frac{V_{surf}^*}{W^{*\frac{3}{2}}}  \approx\frac{4}{15}\left(\frac{35}{4}\right)^{\frac{5}{7}} \left (I_0  \sqrt{ \phi  \cos(\theta_s)}\right )^{\frac{2}{7}} \left(\frac{\Delta\mu}{\mu_w}\right)^{\frac{2}{7}}\left (\frac{Q^*}{W^{*\frac{5}{2}}}\right )^{\frac{5}{7}},
\label{eq:approxvq}
\end{equation}

\begin{equation}
\frac{h^*}{W^*}  \approx \left ( \frac{4}{35}I_0  \sqrt{ \phi  \cos(\theta_s)} \frac{\mu_w}{\Delta\mu}\right )^{-\frac{2}{7}} \left (\frac{Q^*}{W^{*\frac{5}{2}}}\right )^{\frac{2}{7}}.
\label{eq:approxhq}
\end{equation}

\newpage

\end{document}